\documentclass[12pt]{article}
\usepackage{amsmath,amssymb,bm,graphicx}
\usepackage{color} 

\newcommand{\delslash}{\not \! \partial}

\usepackage{hyperref}
\usepackage{cite}

\begin{document}


\begin{center}
{\Large{\bf Majorana Neutrino as Bogoliubov Quasiparticle}}
\end{center}
\vskip .5 truecm
\begin{center}
{\bf { Kazuo Fujikawa$^{1,2}$ and Anca Tureanu$^1$}}
\end{center}

\begin{center}
\vspace*{0.4cm} 
{\it {$^1$Department of Physics, University of Helsinki, P.O.Box 64, 
\\FIN-00014 Helsinki,
Finland\\
$^2$Quantum Hadron Physics Laboratory, RIKEN Nishina Center,\\
Wako 351-0198, Japan
}}
\end{center}
\begin{abstract}
We suggest that the Majorana neutrino should be regarded as a Bogoliubov quasiparticle that is consistently understood only by use of  a relativistic analogue of the Bogoliubov transformation.
The unitary charge conjugation condition ${\cal C}\psi{\cal C}^{\dagger}=\psi$ is not maintained in the definition of a quantum Majorana fermion from a Weyl fermion.  This is remedied by the  Bogoliubov transformation accompanying a redefinition of the charge conjugation properties of vacuum, such that a C-noninvariant fermion number violating term (condensate) is converted to a Dirac mass. We also comment on the chiral symmetry of a Majorana fermion; a
massless Majorana fermion is invariant under a global chiral transformation $\psi\rightarrow \exp[i\alpha\gamma_{5}]\psi$ and different Majorana fermions are distinguished by different chiral $U(1)$ charge assignments.
The reversed process, namely, the definition of 
a Weyl fermion from a well-defined massless Majorana fermion is also briefly discussed.
\end{abstract}
\section{Introduction}
The Majorana fermions received much attention recently not only in particle physics in $d=1+3$~\cite{fukugita, giunti, bilenky, xing} but also in condensed matter physics in $d=1+3$ or less dimensions~\cite{jackiw, beenakker, wilczek}. 
In the massless case, it is generally believed that the Majorana fermion and the Weyl fermion are identical 
in $d=1+3$ as is seen by writing them in the two-component spinor notation. 
Nevertheless, it is not obvious at all how a self-conjugate object is identical to a complex chiral object. 
In this paper, we discuss some basic properties of Majorana and Weyl fermions using  a relativistic analogue of Bogoliubov transformation in $d=1+3$ space-time.  Surprisingly, this analysis leads to the idea that the Majorana neutrino should be regarded as a Bogoliubov quasiparticle that is consistently understood, as it will be explained further, only by use of the Bogoliubov transformation.
The Majorana neutrino could thus become the first Bogoliubov quasiparticle observed in particle physics.
 
\section{Bogoliubov transformation}

It is customary to define a Majorana fermion which satisfies
\begin{eqnarray}\label{matrix_cc}
\psi_{M}(x)=C\overline{\psi_{M}}^{T}(x)
\end{eqnarray}
from a chiral Weyl fermion which satisfies
\begin{eqnarray}
\gamma_{5}\psi_{W}(x)_{R}=\psi_{W}(x)_{R},
\end{eqnarray}
in the manner~\cite{fukugita, giunti, bilenky, xing} 
\begin{eqnarray}\label{naive-Majorana}
\psi_{M}(x)=\psi_{W}(x)_{R}+C\overline{\psi_{W}(x)_{R}}^{T},
\end{eqnarray}
such that $\psi_{M}(x)=C\overline{\psi_{M}}^{T}(x)$. Here, $C$ is the charge conjugation matrix.
Our notational conventions follow those in~\cite{bjorken}. If one starts with a Dirac fermion $\psi_{D}(x)$, we have $\psi_{W}(x)_{R}=\psi_{D}(x)_{R}$  but we do not allow to use $\psi_{W}(x)_{L}=\psi_{D}(x)_{L}$ in addition to $\psi_{W}(x)_{R}$ for a moment. 

In quantum field theory the simple matrix operation \eqref{matrix_cc} has to correspond to the application of a unitary $\cal C$ operator to the quantum fields. In the quantum framework, the definition of a charge conjugated spinor as $\psi^c = C\overline{\psi}^{T}$ can be regarded as a classical operation, for which a quantum realization $\cal C$ has to exist.

To satisfy the operator relation ${\cal C}\psi_{M}(x){\cal C}^{\dagger}=\psi_{M}(x)$, it is often assumed that the charge conjugation is given by~\cite{fukugita, giunti, bilenky, xing} $(\psi_{W}(x)_{R})^{C}= C\overline{\psi_{W}(x)_{R}}^{T}$
or, in the operator notation,
\begin{eqnarray}\label{naive-charge-conjugation}
{\cal C}\psi_{W}(x)_{R}{\cal C}^{\dagger}= C\overline{\psi_{W}(x)_{R}}^{T},
\end{eqnarray}
by presuming a suitable operator ${\cal C}$ defined on an unspecified vacuum. However, this leads to a puzzling result for the unitary charge conjugation operator using $\psi_{W}(x)_{R}=\frac{(1+\gamma_{5})}{2}\psi_{W}(x)_{R}$ and~\cite{FT}:
\begin{eqnarray}\label{contradicting-C}
{\cal C}\psi_{W}(x)_{R}{\cal C}^{\dagger}=\frac{(1+\gamma_{5})}{2}{\cal C}\psi_{W}(x)_{R}{\cal C}^{\dagger}=\frac{(1+\gamma_{5})}{2}C\overline{\psi_{W}(x)_{R}}^{T}=0.
\end{eqnarray}
Moreover, the well-known C- and P-violating weak interaction Lagrangian is written as 
\begin{eqnarray}\label{weak-int}
{\cal L}_{{\rm Weak}}&=&(g/\sqrt{2})\bar{e}_{L}\gamma^{\mu}W^{(-)}_{\mu}(x)\nu_{L}+ h.c.\nonumber\\
&=&(g/\sqrt{2})\bar{e}_{L}\gamma^{\mu}W^{(-)}_{\mu}(x)[(1-\gamma_{5})/2]\nu_{L}+ h.c. 
\end{eqnarray}
If one assumes again ${\cal C}\psi_{W}(x)_{L}{\cal C}^{\dagger}= C\overline{\psi_{W}(x)_{L}}^{T}$ as C-transformation law, one obtains ambiguous results, namely, the first expression implies that ${\cal L}_{{\rm Weak}}$ is invariant under C, while the second expression implies  ${\cal L}_{{\rm Weak}}\rightarrow 0$~\cite{FT}. 
The quantity $(\psi_{W}(x)_{R})^{C}= C\overline{\psi_{W}(x)_{R}}^{T}$ represents a convenient auxiliary object, but not a charge conjugation of $\psi_{W}(x)_{R}$. We are unable to maintain natural operator charge conjugation in \eqref{naive-Majorana} in this construction. 

On the other hand, one may define the charge conjugation by 
\begin{eqnarray}
{\cal C}_{\psi}\psi_{W}(x)_{R}{\cal C}_{\psi}^{\dagger}= C\overline{\psi_{W}(x)_{L}}^{T}
\end{eqnarray}
by noting that $\psi_{W}(x)_{R}=\frac{1+\gamma_{5}}{2}\psi_{D}(x)$ and using the conventional charge conjugation ${\cal C}_{\psi}\psi_{D}(x){\cal C}_{\psi}^{\dagger}= C\overline{\psi_{D}(x)}^{T}$ of a Dirac field $\psi_{D}(x)$. To make the statement precise, we denote the $\cal C$ for $\psi_{D}$ as ${\cal C}_{\psi}$. In this case we do not encounter any obvious contradictions, but we have for the would-be Majorana field in \eqref{naive-Majorana} 
\begin{eqnarray}\label{naive-Majorana-2}
{\cal C}_{\psi}\psi_{M}(x){\cal C}_{\psi}^{\dagger}&=&{\cal C}_{\psi}\psi_{W}(x)_{R}{\cal C}_{\psi}^{\dagger}+{\cal C}_{\psi}C\overline{\psi_{W}(x)_{R}}^{T}{\cal C}_{\psi}^{\dagger}\nonumber\\
&=&C\overline{\psi_{W}(x)_{L}}^{T}+\psi_{W}(x)_{L}\nonumber\\
&\neq&\psi_{M}(x).
\end{eqnarray}
Namely, we can not satisfy the Majorana condition of $\psi_{M}(x)$.
This implies that the subtle aspect of the definition of the operator charge 
conjugation is not solved by the mere change of the convention, but rooted at a more fundamental level. 
Practically,  it is important that this difficulty persists independently of the masses of Weyl and Majorana fermions, for example, in the analysis of the seesaw mechanism.

We can however satisfy consistent operatorial CP,
\begin{eqnarray}
{\cal CP}\psi_{M}(x){\cal (CP)}^{\dagger}=i\gamma^{0}\psi_{M}(t,-\vec{x}),
\end{eqnarray}
which is the fundamental symmetry in a parity violating theory. We adopt the parity operator with the action ${\cal P}\psi(x){\cal P}^{\dagger}=i\gamma^{0}\psi(t,-\vec{x})$ as the natural choice for Majorana fields, since it preserves the reality of the field in the Majorana representation.

As a way to treat the charge conjugation in a transparent manner,  we introduced a relativistic analogue of Bogoliubov transformation, $(\psi, \psi^{c})\rightarrow (N, N^{c})$, defined as
\begin{eqnarray}\label{Bogoliubov}
\left(\begin{array}{c}
            N(x)\\
            N^{c}(x)
            \end{array}\right)
&=& \left(\begin{array}{c}
            \cos\theta\, \psi(x)-\gamma_{5}\sin\theta\, \psi^{c}(x)\\
            \cos\theta\, \psi^{c}(x)+\gamma_{5}\sin\theta\, \psi(x)
            \end{array}\right),
\end{eqnarray}
with a suitable parameter $\theta$~\cite{FT}. Note that $N^{c}=C\bar{N}^{T}$ and $\psi^{c}=C\bar{\psi}^{T}$, and the transformation satisfies  the (classical) consistency condition $N^{c}=C\bar{N}^{T}$ using the expressions given by the right-hand sides.
This Bogoliubov transformation maps a linear combination of a Dirac fermion $\psi$ and its charge conjugate $\psi^{c}$ to another Dirac fermion $N$ and its charge conjugate $N^{c}$, and thus the original Fock vacuum  for $\psi$ is mapped to a new vacuum for $N$, at $t=0$, for example.  It is significant that we use Dirac fermions, either massless or massive, in the definition of the present Bogoliubov transformation.
We can then show that the kinetic terms are invariant
\begin{eqnarray}\label{free-Bogoliubov}
{\cal L}&=&\frac{1}{2}\{\bar{N}i\delslash N + \bar{N^{c}}i\delslash N^{c}\}\nonumber\\
&=&\frac{1}{2}\{\bar{\psi}i\delslash \psi + \bar{\psi^{c}}i\delslash \psi^{c}\}.
\end{eqnarray}
Moreover, using \eqref{Bogoliubov}, we can show that the anticommutators are preserved, i.e.,
\begin{eqnarray}\label{anti-comm}
&& \{N(t,\vec{x}), N^{c}(t,\vec{y})\}=\{\psi(t,\vec{x}), \psi^{c}(t,\vec{y})\},\nonumber\\  
&&\{N_{\alpha}(t,\vec{x}), N_{\beta}(t,\vec{y})\}=\{N^{c}_{\alpha}(t,\vec{x}), N^{c}_{\beta}(t,\vec{y})\}=0.
\end{eqnarray}  
Thus, the canonicity condition of the Bogoliubov transformation is satisfied, irrespective of the masses of the fields $\psi$ and $N$, more generally than that implied by the free fields in \eqref{free-Bogoliubov}. 

It is important that the Bogoliubov transformation \eqref{Bogoliubov} preserves the CP symmetry as a unitary operation on quantum fields,
although it does not preserve the transformation properties under $i\gamma^{0}$-parity  and C separately~\cite{FT}. 
To be precise, the Bogoliubov transformation is a canonical transformation and thus we expect that the dynamical properties on the  vacuum for $\psi$ and the new vacuum for $N(x)$ are mostly equivalent,  but the charge conjugation property is critically changed. 
A transformation analogous to \eqref{Bogoliubov} has been successfully used in the analysis of neutron-antineutron oscillations~\cite{FT1} and the analysis of the hitherto unrecognized sizable fine-tuning~\cite{FT} in the see-saw mechanism~\cite{minkowski,yanagida,mohapatra}.

We now suggest the use of the above Bogoliubov transformation as a means to evade the difficulty associated with the charge conjugation of Weyl and Majorana fermions in a more general context. The transformation \eqref{Bogoliubov} with $\theta=\pi/4$ gives
\begin{eqnarray}\label{Bogoliubov-to-Majorana}
\frac{1}{\sqrt{2}}(N(x)+N^{c}(x))&=& \psi_{R}(x)+C\overline{\psi_{R}}^{T}(x),\nonumber\\
\frac{1}{\sqrt{2}}(N(x)-N^{c}(x))&=& \psi_{L}(x)-C\overline{\psi_{L}}^{T}(x),
\end{eqnarray}
namely, two Majorana fermions, 
\begin{eqnarray}\label{Majorana-fermions}
&&\psi_{M}^{(1)}= \frac{1}{\sqrt{2}}(N(x)+N^{c}(x)),\nonumber\\
&&\psi_{M}^{(2)}=\frac{1}{\sqrt{2}}(N(x)-N^{c}(x)),
\end{eqnarray}
are naturally defined in terms of the new field  $N(x)$ introduced by the Bogoliubov transformation (Bogoliubov quasifermion), with the property:
\begin{eqnarray}
{\cal C}_N N{\cal C}^{\dagger}_N=C\overline{N}^{T}.
\end{eqnarray}
Here we denoted the charge conjugation operator for $N$ as ${\cal C}_{N}$.

Those Majorana fermions satisfy ${\cal C}_{N}\psi_{M}^{(1)}{\cal C}_{N}^{\dagger}=\psi_{M}^{(1)}$ and $\psi_{M}^{(1)}=C\overline{\psi_{M}^{(1)}}^{T}$, and ${\cal C}_{N}\psi_{M}^{(2)}{\cal C}_{N}^{\dagger}= -\psi_{M}^{(2)}$ and $\psi_{M}^{(2)}=-C\overline{\psi_{M}^{(2)}}^{T}$, the first being even and the second odd eigenfield of the charge conjugation operator ${\cal C}_{N}$.
The fields $\psi_{M}^{(1)}$ and $\psi_{M}^{(2)}$  correspond to the conventional definitions of Majorana fermions in terms of Weyl fermions on the right-hand side of \eqref{Bogoliubov-to-Majorana}, that do not support the operator charge conjugation ${\cal C}_{\psi}$.  Incidentally, the definition of Majorana fermions by itself implies a certain "condensation" of the fermion number in the vacuum (see Ref. \cite{chang}). 
The Bogoliubov transformation helps define the eigenstates of the charge conjugation operator ${\cal C}_{N}$ in a consistent manner.

  In the case of massless Majorana and  Weyl fermions, we do not encounter an explicit fermion number violation in the Lagrangian, for example,
\begin{eqnarray}\label{free-Majorana-Weyl}
{\cal L}&=&\frac{1}{2}\overline{\psi_{M}^{(1)}(x)}i\delslash\psi_{M}^{(1)}(x)\nonumber\\
&=&\overline{\psi_{R}(x)}i\delslash \psi_{R}(x),
\end{eqnarray}
unlike the Bardeen--Cooper--Schrieffer (BCS) theory~\cite{jackiw, beenakker} or the see-saw mechanism~\cite{minkowski,yanagida,mohapatra}, where there is an energy or mass gap. Yet the chiral Weyl fermion $\psi_{R}(x)$, in its strict definition, is not
the eigenstate of charge conjugation $\cal C$ and parity $\cal P$ transformations, although ${\cal CP}$ is well-defined.
Thus the definition of the exact eigenstate of ${\cal C}_{\psi}$, namely the free Majorana fermion,
has certain conflicts with the definition of the charge conjugation for $\psi_{R}(x)$.\footnote
{One may recall that the free Lagrangian ${\cal L}=\overline{\psi_{R}(x)}i\delslash \psi_{R}(x)=\overline{\psi_{R}(x)}i\delslash [(1+\gamma_{5})/2]\psi_{R}(x)$ suffers from the ambiguity if one adopts the charge conjugation  in \eqref{naive-charge-conjugation}, although we have already rejected $(\psi_{R}(x))^{C}= C\overline{\psi_{R}(x)}^{T}$ as a transformation rule of charge conjugation.}  
These conflicts are resolved by  the charge conjugation ${\cal C}_{N}$ after the Bogoliubov transformation, which is precisely what the first relation of \eqref{Bogoliubov-to-Majorana} implies.
We however implicitly assumed the exsitence of a Dirac fermion in defining the Bogoliubov transformation, which needs to be remembered when we consider applications. 

We can also solve \eqref{Bogoliubov} with $\theta=\pi/4$ in terms of the Majorana fermions \eqref{Majorana-fermions} as 
\begin{eqnarray}\label{Majorana-to-Dirac}
\left(\begin{array}{c}
            \psi(x)\\
            \psi^{c}(x)
            \end{array}\right)
&=& \left(\begin{array}{c}
            \left(\frac{1+\gamma_{5}}{2}\right)\psi_{M}^{(1)}(x)+\left(\frac{1-\gamma_{5}}{2}\right)     
            \psi_{M}^{(2)}(x)\\
            \left(\frac{1-\gamma_{5}}{2}\right)\psi_{M}^{(1)}(x)-\left(\frac{1+\gamma_{5}}{2}\right)     
            \psi_{M}^{(2)}(x)            
            \end{array}\right).
\end{eqnarray}
The Majorana fermions $\psi_{M}^{(1)}(x)$ and $\psi_{M}^{(2)}(x)$, if chosen as the primary dynamical degrees of freedom, belong to definite representations of the basic symmetries P and T and thus C, due to the CPT symmetry of field theory on the Minkowski space-time. With this choice of fundamental fields, the natural quantum realization of the charge conjugation in \eqref{Majorana-to-Dirac} is ${\cal C}_N$, under which ${\cal C}_{N}\psi_{M}^{(1)}(x){\cal C}_{N}^{\dagger}\rightarrow \psi_{M}^{(1)}(x)$ and ${\cal C}_{N}\psi_{M}^{(2)}(x){\cal C}_{N}^{\dagger}\rightarrow -\psi_{M}^{(2)}(x)$. However, on the left-hand side of \eqref{Majorana-to-Dirac} this operation does not send $\psi$ to $\psi^c$, which would be expected if the operator charge conjugation is preserved. On the other hand, the classical consistency condition  $\psi^{c}(x)=C\overline{\psi(x)}^{T}$ is satisfied. This inconsistency is precisely the difficulty we encountered in the construction of the Majorana fermions {\em via Weyl fermions} in \eqref{naive-Majorana}. The Majorana fermions $\psi_{M}^{(1,2)}(x)$ are consistently defined in terms of the Bogoliubov $N(x)$ and $N^{c}(x)$, but not consistenly in terms of the chiral projected components $\psi_{L,R}(x)$. 

As for the physical implication of the above analysis, 
 it may be natural to accept the description of the Majorana neutrino using the Bogoliubov transformation as a physical one in the Standard Model, since we start with the chiral Weyl fermions as the basic building blocks of  gauge theory. As for the Dirac fermion $\psi(x)$, which plays an important role in \eqref{Bogoliubov-to-Majorana}, it is effectively produced in the see-saw mechanism by the addition of the right-handed neutrino. One may thus regard the possible Majorana neutrino as a first Bogoliubov quasiparticle, which is consistently understood only by the use of the Bogoliubov transformation.

\section{Chiral symmetry and related issues}

We now come back to the analysis of mostly massless fermions.
From the point of view of conserved symmetries, the massless Dirac fermion has the $U(1)$ fermion number and chiral $U(1)$ symmetries. The Weyl fermion picks up the chiral choices $(1\pm \gamma_{5})/2$ as the conserved symmetry. The massless Majorana fermion retains only the chiral $\gamma_{5}$ symmetry of the Dirac fermion as a conserved symmetry, as is seen by the construction $\psi_{M}(x)=(1/2)[\psi_{D}(x)+C\overline{\psi_{D}}^{T}(x)]$, which implies 
$\psi_{D}\rightarrow e^{i\alpha\gamma_{5}}\psi_{D}\ \Rightarrow \  \psi_{M}\rightarrow e^{i\alpha\gamma_{5}}\psi_{M}$,
or more directly 
\begin{eqnarray}
e^{i\alpha\gamma_{5}}\psi_{M}=C\overline{e^{i\alpha\gamma_{5}}\psi_{M}}^{T}
\end{eqnarray}
if one uses $\psi_{M}=C\overline{\psi_{M}}^{T}$. The quantity $i\gamma_{5}$ is real in the Majorana representation of $\gamma$ matrices.
If one insists on the eigenstates of the chiral symmetry, the chiral fields $\psi_{M}(x)_{L,R}$ with $\psi_{M}(x)=\psi_{M}(x)_{L}+\psi_{M}(x)_{R}$ are picked up.

The chiral symmetry implies the invariance under 
\begin{eqnarray}
\psi_{M}(x)_{L}\rightarrow e^{-i\alpha}\psi_{M}(x)_{L},\ \ \ \psi_{M}(x)_{R}\rightarrow e^{i\alpha}\psi_{M}(x)_{R}.
\end{eqnarray}
It is important that we have the same $\alpha$ in $e^{\pm i\alpha}\psi_{M}(x)_{L,R}$. In contrast, we have two parameters $\alpha$ and $\beta$ in the case of a Dirac fermion:
\begin{eqnarray}\label{Dirac-chiral-symmetry}
\psi_{D}(x)_{L}\rightarrow e^{-i\alpha}\psi_{D}(x)_{L},\ \ \ \psi_{D}(x)_{R}\rightarrow e^{-i\beta}\psi_{D}(x)_{R},
\end{eqnarray}
but for a Weyl fermion $\psi_{W}(x)_{L}=\psi_{D}(x)_{L}$, we have obviously only a single parameter.
When one regards the quantities defined from Weyl fermions 
\begin{eqnarray}
&&\psi^{(1)}_{M}(x)=\psi_{W}(x)_{R}+C\overline{\psi_{W}(x)}^{T}_{R},\nonumber\\
&&\psi^{(2)}_{M}(x)=\psi_{W}(x)_{L}-C\overline{\psi_{W}(x)}^{T}_{L},
\end{eqnarray}
as Majorana fermions, the distinct chiral symmetries of these Majorana fermions arise from those two different chiral symmetries. 

In general, each massless Majorana fermion has its own chiral symmetry generated by the generic chiral transformation $e^{i\alpha\gamma_{5}}$, and different Majorana fermions are distinguished by the different charge assignments to these $U(1)$ chiral transformations.

As for the neutrinoless double beta decay, one may consider
\begin{eqnarray}\label{beta-decay}
{\cal L}= (g/\sqrt{2})\bar{e}_{L}\gamma^{\mu}W_{\mu}\frac{(1-\gamma_{5})}{2}\psi_{M} + h.c.,
\end{eqnarray}
which projects the Majorana neutrino to the chiral state $\psi_{M}(x)_{L}$. 
It is well-known that the neutrinoless double beta decay takes place only with massive Majorana neutrinos~\cite{schechter}. We here briefly comment on this criterion from the point of view of the chiral symmetry
breaking by the mass term. The chiral symmetry breaking may be characterized by 
\begin{eqnarray}
\langle 0| \overline{\psi(x)}\psi(x)|0\rangle=\lim_{x\rightarrow y}\langle 0|T^{\star} \overline{\psi(x)}\psi(y)|0\rangle\neq 0,
\end{eqnarray}
although we are dealing with explicit chiral symmetry breaking. The use of 
$\langle 0| \overline{\psi(x)}\psi(x)|0\rangle$ as an indicator of chiral symmetry is
well-known~\cite{Nambu-Jona-Lasinio, Gell-Mann}. For a Majorana 
fermion, $\psi(x)=\nu_{L}(x)+\nu_{R}(x)$, it is confirmed that this is equivalent to 
\begin{eqnarray}\label{criterion-double-beta-decay}
\lim_{x\rightarrow y}\langle 0|T^{\star} \nu^{T}_{L}(x)C\nu_{L}(y)|0\rangle\neq 0
\end{eqnarray}
by noting $\overline{\nu_{R}}=\nu_{L}^{T}C $,
with a suitable regularization such as the dimensional regularization, where
$C$ is the charge conjugation matrix.
Chiral symmetry for either a massless Majorana neutrino or a Weyl neutrino predicts the vanishing result of this correlation to be consistent with the conclusion in~\cite{schechter}. As for the Majorana propagator generated by the neutrinoless double beta decay amplitude discussed in~\cite{schechter}, one may combine two ends of the neutrino line of the propagator at a point and  then recognize the consistency with the criterion \eqref{criterion-double-beta-decay}.

As for the extra CP violating phases in the case of massive Majorana neutrinos~\cite{bilenky2, takasugi, schechter2}, they are eliminated by the chiral freedom in the case of massless Majorana neutrinos.   

We finally mention a reversed process, namely, the definition of a Weyl fermion from a massless Majorana fermion that satisfies ${\cal C}\psi_{M}(x){\cal C}^{\dagger}=\psi_{M}(x)$ as well as $\psi_{M}(x)=C\overline{\psi_{M}}^{T}(x)$.
We emphasize that the Majorana fermion defined in \eqref{naive-Majorana} does not satisfy both these conditions. In this scheme, we have
\begin{eqnarray}
\psi_{M}(x)&=&\psi_{M}(x)_{R}+ \psi_{M}(x)_{L}\nonumber\\
&=&\psi_{M}(x)_{R}+ C\overline{\psi_{M}(x)}_{R}^{T}\nonumber\\
&=&\psi_{M}(x)_{L}+ C\overline{\psi_{M}(x)}_{L}^{T},
\end{eqnarray}
with the constraint $\psi_{M}(x)_{L}=C\overline{\psi_{M}(x)}_{R}^{T}$.
A salient feature of the present scheme is that chiral components satisfy the operator charge conjugation properties
\begin{eqnarray}
{\cal C}\psi_{M}(x)_{R}{\cal C}^{\dagger}=\psi_{M}(x)_{R}, \ \ 
{\cal C}\psi_{M}(x)_{L}{\cal C}^{\dagger}=\psi_{M}(x)_{L},
\end{eqnarray}
and one can avoid the inconsistency we encountered in \eqref{naive-Majorana}, since in the present case the Majorana field is considered elementary.
One can thus use one of $\psi_{M}(x)_{L}$, $\psi_{M}(x)_{R}$ and $\psi_{M}(x)=C\overline{\psi_{M}}^{T}(x)$ as a primary dynamical degree of freedom,
\begin{eqnarray}\label{Majorana-Weyl}
{\cal L}_{M}&=&\frac{1}{2}\overline{\psi_{M}(x)}i\delslash \psi_{M}(x)\nonumber\\
&=&\overline{\psi_{M}(x)_{R}}i\delslash \psi_{M}(x)_{R}\nonumber\\
&=&\overline{\psi_{M}(x)_{L}}i\delslash \psi_{M}(x)_{L}.
\end{eqnarray}
One may interpret the superposition  $\psi_{M}(x)=\psi_{M}(x)_{R}+ \psi_{M}(x)_{L}$ in a manner  analogous to the case of the photon polarizations. The interaction chooses one of these possibilities.
The field $\psi_{M}(x)$ which is transformed under parity as
\begin{eqnarray}
{\cal P}\psi_{M}(x){\cal P}^{\dagger}&=&i\gamma^{0}\psi_{M}(t,-\vec{x}),
\end{eqnarray}
is an analogue of linear polarization, but only one linear polarization  (Majorana fermion) appears in the present case.
The variables $\psi_{M}(x)_{L}$ and $\psi_{M}(x)_{R}$ in $\psi_{M}(x)=\psi_{M}(x)_{L}+\psi_{M}(x)_{R}$ are interchanged by parity
\begin{eqnarray}\label{Majorana-P}
{\cal P}\psi_{M}(x)_{R}{\cal P}^{\dagger}&=&i\gamma^{0}\psi_{M}(t,-\vec{x})_{L},\nonumber\\
{\cal P}\psi_{M}(x)_{L}{\cal P}^{\dagger}&=&i\gamma^{0}\psi_{M}(t,-\vec{x})_{R},
\end{eqnarray}
and they are the analogues of the circular polarization but only one of them is allowed at a time, either left-handed or right-handed (Weyl fermions). The Majorana fermion has no {\em a priori} preference for a left- or right-handed state, and the measurement (interaction) will pick up one of $\psi_{M}(x)_{R}$, $\psi_{M}(x)_{L}$ or $\psi_{M}(x)=\psi_{M}(x)_{R}+ \psi_{M}(x)_{L}$.
This subject, besides an interest in the quantum information carried by ultra-relativistic particles~\cite{kf}, may become relevant in the future if one takes the Majorana fermion as a fundamental entity of Nature.
\\

In conclusion, we have discussed the basic properties of Majorana and Weyl fermions and the relativistic analogue of Bogoliubov transformation in $d=1+3$. The present study of the Majorana fermion as a Bogoliubov quasiparticle may be compared to  condensed matter physics where the Bogoliubov quasiparticle is well-known but the notion of Majorana particles is new~\cite{jackiw, beenakker,wilczek}. In the Standard Model it is natural to start with Weyl fermions, and we argued that a change of vacuum as is indicated by the Bogoliubov transformation is inevitable to understand the possible Majorana neutrino consistently. 
The Majorana neutrino is then regarded as a Bogoliubov quasiparticle for the first time in particle physics.
\\

We thank Masud Chaichian for very helpful discussions. This work is supported in part by the Magnus Ehrnrooth Foundation. The support of the Academy of Finland under the Projects no. 136539 and 272919 is gratefully acknowledged.

\appendix

\section{Appendix: Explicit example}
We now illustrate the idea stated above, including the definitions of the charge conjugation operator and vacuum, using an explicit example of the single flavor seesaw model defined by \cite{FT} 
\begin{eqnarray}\label{3}
{\cal L}&=&(1/2)\{\overline{\nu}(x)[i\delslash - m_{D}]\nu(x)+ \overline{\nu^{c}}(x)[i\delslash - m_{D}]\nu^{c}(x)\}\nonumber\\
&-&(\epsilon_{1}/4)[\overline{\nu^{c}}(x)\nu(x) +\overline{\nu}(x)\nu^{c}(x)]\nonumber\\
&-&(\epsilon_{5}/4)[\overline{\nu^{c}}(x)\gamma_{5}\nu(x) -\overline{\nu}(x)\gamma_{5}\nu^{c}(x)],
\end{eqnarray}
where  we used a Dirac-type variable
\begin{eqnarray}\label{2}
\nu(x)\equiv \nu_{L}(x) + n_{R}(x)
\end{eqnarray}
and $\epsilon_{1}=m_{R}+m_{L}$ and $\epsilon_{5}=m_{R}-m_{L}$, which are real  if one assumes CP symmetry, for simplicity.
The above  Lagrangian \eqref{3} is CP conserving, although C, $\nu\rightarrow C\overline{\nu}^{T}$, and P ($i\gamma^{0}$-parity) are separately broken by the last term. 

Let us first recapitulate the traditional approach to the single flavour seesaw and point out the issues with charge conjugation, and subsequently present on the same example our proposed approach using the Bogoliubov transformation and indicate how the issue is solved. Usually, one exactly diagonalizes the Lagrangian \eqref{3} (see, for example, Ref. \cite{bilenky}) as 
\begin{eqnarray}\label{exact-solution1}
{\cal L}
&=&\overline{\tilde{\nu}}(x)i\delslash \tilde{\nu}(x)\nonumber\\
&-&(1/2)\left(
 \overline{\tilde{\nu}_{R}}M_{1}\tilde{\nu}_{L}^{c}-\overline{\tilde{\nu}_{R}^{c}}M_{2}\tilde{\nu}_{L}\right) +h.c.,
\end{eqnarray}
where
\begin{eqnarray} \label{variable-change}          
            &&\left(\begin{array}{c}
            \tilde{\nu}_{L}^{c}\\
            \tilde{\nu}_{L}
            \end{array}\right)\equiv O            
            \left(\begin{array}{c}
            \nu_{L}^{c}\\
            \nu_{L}
            \end{array}\right)
            ,\nonumber\\ 
            &&\left(\begin{array}{c}
            \tilde{\nu}_{R}\\
            \tilde{\nu}_{R}^{c}
            \end{array}\right)\equiv O            
            \left(\begin{array}{c}
            \nu_{R}\\
            \nu_{R}^{c}
            \end{array}\right),  
\end{eqnarray}
with a suitable $2\times 2$ orthogonal matrix $O$. The mass matrix in \eqref{3} is diagonalized as 
\begin{eqnarray}
            O
            \left(\begin{array}{cc}
            \frac{1}{2}(\epsilon_{1}+\epsilon_{5})& m_{D}\\
            m_{D}&\frac{1}{2}(\epsilon_{1}-\epsilon_{5})
            \end{array}\right)
            O^{T}
            =\left(\begin{array}{cc}
            M_{1}&0\\
            0&-M_{2}
            \end{array}\right)    ,        
\end{eqnarray}
where  
\begin{equation}\label{m_exact_diag}
M_{1,2}=\sqrt{m^{2}+(\epsilon_{5}/2)^{2}} \pm \epsilon_{1}/2
\end{equation}  
are real eigenvalues. 
In our convention, $\nu_{L}^{c}=[(1-\gamma_{5})/2]\nu^{c}$ and $\nu_{R}^{c}=[(1+\gamma_{5})/2]\nu^{c}$ are left- and right-handed, respectively.
Then one defines Majorana-type fields by
\begin{eqnarray}\label{Majorana-fields2}
\tilde{\psi}_{+}(x)&=&\tilde{\nu}_{R}(x)+\tilde{\nu}_{L}^{c}(x),\nonumber\\
\tilde{\psi}_{-}(x)&=&\tilde{\nu}_{L}(x)-\tilde{\nu}_{R}^{c}(x),
\end{eqnarray}
and the Lagrangian becomes
\begin{eqnarray}\label{Majorana2}
 {\cal L}
&=&\frac{1}{2}\overline{\tilde{\psi}_{+}}(x)[i\delslash - M_{+}]\tilde{\psi}_{+}(x)
+\frac{1}{2}\overline{\tilde{\psi}_{-}}(x)[i\delslash - M_{-}]\tilde{\psi}_{-}(x).            
\end{eqnarray}
where we denoted $M_{1}=M_{+}$ and $M_{2}=M_{-}$, respectively. The mass $M_-$ represents the tiny neutrino mass.

However, one cannot show that the fields $\tilde{\psi}_{+}$ and $\tilde{\psi}_{-}$ are truly Majorana in the quantum field theory sense, just as we failed to show that $\psi_M$ in \eqref{naive-Majorana} is a Majorana field.
This analysis shows that one can diagonalize the C-violating Lagrangian exactly in terms of Weyl fermions as in \eqref{exact-solution1}, but it is  not invariant operatorially under C. Thus one cannot rewrite this C-violating Lagrangian in terms of Majorana fermions which are exact eigenstates of a quantum {\cal C}-operator. This is the origin of the puzzling aspects we encountered in \eqref{naive-Majorana}. This feature is also shared with the simple case in \eqref{free-Majorana-Weyl}.

To evade this problem, we recall that charge conjugation is well-defined for a Dirac field and that the Majorana field is supposed to be exactly invariant under charge conjugation. Therefore we propose to consider a different path, using the Bogoliubov transformation  of Dirac fields as in \eqref{10}.
Thus the appearance of the Bogoliubov transformation is generic for the C-violating Lagrangian in the seesaw mechanism. \footnote{One the other hand, if one sets $\epsilon_{5}=0$ in the Lagrangian \eqref{3}, the model becomes C-invariant although it still violates the fermion number. In such a case, one can directly rewrite the Lagrangian in terms of Majorana fermions in a consistent manner without the help of the Bogoliubov transformation. (This case corresponds to the model analyzed in \cite{chang}).}

Returning to the Lagrangian \eqref{3}, we apply the Bogoliubov transformation $(\nu, \nu^{c})\rightarrow (N, N^{c})$, defined as
\begin{eqnarray}\label{10}
\left(\begin{array}{c}
            N(x)\\
            N^{c}(x)
            \end{array}\right)
&=& \left(\begin{array}{c}
            \cos\theta\, \nu(x)-\gamma_{5}\sin\theta\, \nu^{c}(x)\\
            \cos\theta\, \nu^{c}(x)+\gamma_{5}\sin\theta\, \nu(x)
            \end{array}\right),
\end{eqnarray}
with
\begin{eqnarray}\label{mixing-angle}
\sin 2\theta =(\epsilon_{5}/2)/\sqrt{m^{2}+(\epsilon_{5}/2)^{2}}.
\end{eqnarray}
We can then show that the anticommutators are preserved, i.e.,
\begin{eqnarray}\label{anti-comm}
&& \{N(t,\vec{x}), N^{c}(t,\vec{y})\}=\{\nu(t,\vec{x}), \nu^{c}(t,\vec{y})\},\nonumber\\  
&&\{N_{\alpha}(t,\vec{x}), N_{\beta}(t,\vec{y})\}=\{N^{c}_{\alpha}(t,\vec{x}), N^{c}_{\beta}(t,\vec{y})\}=0,
\end{eqnarray}  
hence the transformation is canonical.

After the Bogoliubov transformation, which diagonalizes the Lagrangian with $\epsilon_{1}=0$,  ${\cal L}$ in \eqref{3} becomes
\begin{eqnarray}\label{13}
{\cal L}&=&\frac{1}{2}\left[\overline{N}(x)\left(i\delslash - M\right)
 N(x)+\overline{N^{c}}(x)\left(i\delslash - M\right)N^{c}(x)\right]\nonumber\\
&-&\frac{\epsilon_{1}}{4}[\overline{N^{c}}(x)N(x) + \overline{N}(x)N^{c}(x)],
\end{eqnarray}
with the mass parameter
\begin{eqnarray}\label{14}
M\equiv \sqrt{m^{2}+(\epsilon_{5}/2)^{2}}.
\end{eqnarray}
This implies that the Bogoliubov transformation maps the original theory to
a theory characterized by the new (large) mass scale $M$, and $\epsilon_{5}/2$ corresponds to the energy gap, in analogy with the BCS theory. 
We emphasize that the Bogoliubov transformation \eqref{10} preserves the CP symmetry, although it does not preserve the transformation properties under $i\gamma^{0}$-parity and C separately.
In the present single flavor model, this leads to the Lagrangian \eqref{13} of the  Bogoliubov quasi-fermion $N(x)$ which is symmetric under the $i\gamma^{0}$-parity and C transformations, $N(t,\vec{x})\rightarrow i\gamma^{0}N(t,-\vec{x})$ and $N(x)\rightarrow C\overline{N}^{T}$, respectively. 

The combinations
\begin{eqnarray}\label{Majorana-fields1} 
\psi_{+}(x)=\frac{1}{\sqrt{2}}(N(x)+N^{c}(x)),\ \ \ 
\psi_{-}(x)=\frac{1}{\sqrt{2}}(N(x)-N^{c}(x))
\end{eqnarray}   
represent Majorana fields if one uses the charge conjugation operator defined by $N(x)\rightarrow N^{c}(x)$, and
the Lagrangian \eqref{13} is exactly diagonalized in the form
\begin{eqnarray}\label{Majorana1}
{\cal L}&=&\frac{1}{2}\{\overline{\psi}_{+}[i\delslash-M_{+}]\psi_{+}
+\overline{\psi}_{-}[i\delslash-M_{-}]\psi_{-}\},
\end{eqnarray}
with the masses 
\begin{eqnarray}\label{masses}
M_{\pm}=M \pm \epsilon_{1}/2\equiv \sqrt{m^{2}+(\epsilon_{5}/2)^{2}}\pm \epsilon_{1}/2,
\end{eqnarray}
such that $m_{\nu}=M_{-}$ corresponds to the small neutrino mass. As expected, the mass eigenvalues coincide with those obtained by the previously presented direct diagonalization of the mass matrix, eq. \eqref{m_exact_diag}.

As for the explicit definitions of the charge conjugation operators and vacua, one may ideally define them directly. However, the direct definitions turn out to be very involved. We thus adopt the following consistency argument which can be formulated rigorously.

We recall that the Lagrangians \eqref{Majorana1} and \eqref{Majorana2} contain fields with the same masses and classically  expected  to be Majorana fermions, $\psi_{+}^{c}=\psi_{+}$ and $\psi_{-}^{c}=-\psi_{-}$, respectively. They satisfy the free equations for fermions:
\begin{eqnarray}\label{Majorana_eom}
&&[i\delslash - M_{+}]\psi_{+}(x)=0,\nonumber\\
&&[i\delslash - M_{-}]\psi_{-}(x)=0.            
\end{eqnarray}
These free Dirac equations are solved exactly, and the vacuum is defined by
\begin{eqnarray}\label{Vacuum}
\psi_{+}^{(+)}(x)|0\rangle_{M}=\psi_{-}^{(+)}(x)|0\rangle_{M} =0       ,
\end{eqnarray}
where $\psi_{\pm}^{(+)}(x)$ stand for positive frequency components.
One can also construct the operator charge conjugation ${\cal C}_{M}$ which satisfies 
\begin{eqnarray}\label{Majorana_CC}
{\cal C}_{M}\psi_{+}(x){\cal C}^{\dagger}_{M}=C\overline{\psi_{+}(x)}^{T}=\psi_{+}(x),\ \ \ \
{\cal C}_{M}\psi_{-}(x){\cal C}^{\dagger}_{M}=C\overline{\psi_{-}(x)}^{T}=-\psi_{-}(x),        
\end{eqnarray}
with ${\cal C}_{M}|0\rangle_{M}=|0\rangle_{M}$ by following the procedure in the textbook \cite{bjorken}.  

We next invert \eqref{Majorana-fields1} and \eqref{Majorana-fields2} in the form
\begin{eqnarray}\label{conversion1}
\left(\begin{array}{c}
            N(x)\\
            N^{c}(x)
            \end{array}\right)
&=& \left(\begin{array}{c}
           \frac{1}{\sqrt{2}} (\psi_{+}(x) + \psi_{-}(x))\\
            \frac{1}{\sqrt{2}} (\psi_{+}(x) - \psi_{-}(x))            
            \end{array}\right).
\end{eqnarray}
and 
\begin{eqnarray}\label{conversion2}
\left(\begin{array}{c}
            \tilde{\nu}(x)\\
            \tilde{\nu}^{c}(x)
            \end{array}\right)
&=& \left(\begin{array}{c}
            \left(\frac{1+\gamma_{5}}{2}\right)\psi_{+}(x) + \left(\frac{1-\gamma_{5}}{2}\right)\psi_{-}(x)\\
            \left(\frac{1-\gamma_{5}}{2}\right)\psi_{+}(x) - \left(\frac{1+\gamma_{5}}{2}\right)     
            \psi_{-}(x)           
            \end{array}\right),
\end{eqnarray}    
respectively. 
Both of these relations satisfy the classical charge conjugation conditions, $N^{c}=C\overline{N}^{T}$ and $\tilde{\nu}^{c}=C\overline{\tilde{\nu}}^{T}$, using the expressions on the right-hand sides with $C\overline{\psi_{+}(x)}^{T}=\psi_{+}(x)$ and $C\overline{\psi_{-}(x)}^{T}=-\psi_{-}(x)$.

We satisfy
\begin{eqnarray}
{\cal C}_{M}N(x){\cal C}^{\dagger}_{M}=\frac{1}{\sqrt{2}} {\cal C}_{M}\left(\psi_{+}(x) + \psi_{-}(x)\right){\cal C}^{\dagger}_{M}=N^{c}(x),
\end{eqnarray}
but
\begin{eqnarray}
{\cal C}_{M}\tilde{\nu}(x){\cal C}^{\dagger}_{M}
&=&{\cal C}_{M}\left(  \left(\frac{1+\gamma_{5}}{2}\right)\psi_{+}(x) + \left(\frac{1-\gamma_{5}}{2}\right)\psi_{-}(x) \right){\cal C}^{\dagger}_{M}\nonumber\\
&=&\left(\frac{1+\gamma_{5}}{2}\right)\psi_{+}(x) - \left(\frac{1-\gamma_{5}}{2}\right) \psi_{-}(x)  \nonumber\\
&\neq& \tilde{\nu}^{c}(x).
\end{eqnarray}
One can thus consistently define ${\cal C}_{N}={\cal C}_{M}$ and $|0\rangle_{N}=|0\rangle_{M}$ for the Bogoliubov quasiparticle $N(x)$, but the exact charge conjugation ${\cal C}_{M}$ for the exact solutions of the Majorana fermion does not induce the required charge conjugation of the variable $\tilde{\nu}(x)$, namely, ${\cal C}_{\tilde{\nu}}\neq {\cal C}_{M}$, which in turn implies that the vacuum defined by ${\cal C}_{\tilde{\nu}}|0\rangle_{\tilde{\nu}}=|0\rangle_{\tilde{\nu}}$ differs from $|0\rangle_{M}$, namely, $|0\rangle_{\tilde{\nu}}\neq |0\rangle_{N}$.
The exact solutions of our Lagrangian \eqref{3} are given by two Majorana fermions with different masses, and the true vacuum 
$|0\rangle_{M}$ is defined by a direct product of the vacua for those two Majorana fermions. (The true vacuum 
$|0\rangle_{M}$ differs from the vacuum of $N(x)$ expected from the Dirac part of the Lagrangian \eqref{13} with mass $M$ or the vacuum of $\tilde{\nu}(x)$ expected from the Dirac part of the Lagrangian \eqref{exact-solution1} with a vanishing mass.)  All the  Dirac-type variables $N(x)$ and $\tilde{\nu}$ are expanded in terms of these two Majorana fermions, but the variable $\tilde{\nu}$, which contains chiral projection operators
in the expansion,  cannot be a representation of the charge conjugation operator ${\cal C}_{M}$.

Thus, the Bogoliubov transformation converts the ``C-violating condensate'' with the coefficient $\epsilon_{5}$ in the fermion number violating condensates  in \eqref{3} into a Dirac mass $M=\sqrt{m_{D}^{2}+(\epsilon_{5}/2)^{2}}$. We regard this to be an  analogue of the absorption of the chiral condensate by the massless Dirac fermion to become a massive Dirac fermion in the Nambu--Jona-Lasinio model ~\cite{Nambu-Jona-Lasinio}, with the corresponding changes of vacuum states.  
For the very special case $\epsilon_{5}=0$ and thus the mixing angle $\theta=0$ in \eqref{mixing-angle}, which is irrelevant for the seesaw mechanism but relevant for the neutron oscillations \cite{chang, FT1}, the Lagrangian \eqref{3} becomes C-invariant and the Bogoliubov transformation is not required.  In this case,  effectively $N(x)=\nu(x)$, where $\nu(x)$ is the very original variable in \eqref{3}; one can confirm that 
$\tilde{\psi}_{+}(x)=\tilde{\nu}_{R}(x)+\tilde{\nu}^{c}_{L}(x)=(\nu(x)+\nu^{c}(x))/\sqrt{2}$
and 
$\tilde{\psi}_{-}(x)=\tilde{\nu}_{L}(x)-\tilde{\nu}^{c}_{R}(x)=(\nu(x)-\nu^{c}(x))/\sqrt{2}$
in \eqref{variable-change}. In this special case with C conservation, we thus have ${\cal C}_{\nu}={\cal C}_{M}$, with 
${\cal C}_{\nu}\nu(x){\cal C}^{\dagger}_{\nu}=C\overline{\nu(x)}^{T}=\nu^{c}(x)$, and the true vacuum is given by $|0\rangle_{M}$.

\end{document}